\documentclass[conference]{IEEEtran}
\pdfoutput=1
\usepackage{tikz}

\let\labelindent\relax
\usepackage{comment}
\usepackage{framed,graphicx}
\usepackage{multirow}
\usepackage{rotating}
\usepackage{amsmath}
\usepackage{bigstrut}
\usepackage{color}
\usepackage{silence}
\usepackage{tabularx}
\usepackage{listings}
\usepackage{float}
\usepackage{graphics}
\usepackage{eqparbox}
\usepackage{graphics}
\usepackage{scrextend}
\usepackage{colortbl}
\usepackage{caption}
\usepackage{mathptmx} 
\usepackage[scaled=.90]{helvet} 
\usepackage{courier}
\usepackage{balance}
\usepackage{picture}
\usepackage{algorithm}
\usepackage{algorithmicx}
\usepackage{algpseudocode}
\usepackage[export]{adjustbox}
\renewcommand{\footnotesize}{\scriptsize}
\definecolor{lightgray}{gray}{0.8}
\definecolor{darkgray}{gray}{0.6}
\definecolor{lavenderpink}{rgb}{0.98, 0.68, 0.82}
\definecolor{celadon}{rgb}{0.67, 0.88, 0.69}

\newcommand{\quart}[4]{\begin{picture}(80,4)
	{\color{black}\put(#3,2){\circle*{4}}\put(#1,2){\line(1,0){#2}}}\end{picture}}
\definecolor{Gray}{gray}{0.95}
\definecolor{LightGray}{gray}{0.975}
\definecolor{cellgrey}{rgb}{.87, .87, .87}

\usepackage[lofdepth,lotdepth]{subfig}

\newcommand{\bi}{\begin{itemize}}
	\newcommand{\ei}{\end{itemize}}
\newcommand{\be}{\begin{enumerate}}
	\newcommand{\ee}{\end{enumerate}}
\newcommand{\tion}[1]{\S\ref{sect:#1}}
\newcommand{\fig}[1]{Figure~\ref{fig:#1}}

\newcommand{\eq}[1]{Equation~\ref{eq:#1}}

\usepackage[shortlabels]{enumitem}
\usepackage{url}

\definecolor{steel}{rgb}{.11, .11, .7}
\definecolor{Gray}{rgb}{0.88,1,1}
\definecolor{Gray}{gray}{0.85}
\usepackage[framed]{ntheorem}
\usepackage{framed}
\usepackage{tikz}
\usetikzlibrary{shadows}
\theoremclass{Lesson}
\theoremstyle{break}

\tikzstyle{thmbox} = [rectangle, rounded corners, draw=black,
fill=Gray!1,  drop shadow={fill=black, opacity=1}]

\definecolor{shadecolor}{gray}{1}

\newshadedtheorem{question}{Research Question}

\definecolor{shadecolor}{gray}{0.95}

\newshadedtheorem{lesson}{Research Answer}

\DeclareFixedFont{\ttb}{T1}{txtt}{bx}{n}{12} 
\DeclareFixedFont{\ttm}{T1}{txtt}{m}{n}{12}  

\usepackage{color}
\definecolor{deepblue}{rgb}{0,0,0.5}
\definecolor{deepred}{rgb}{0.6,0,0}
\definecolor{deepgreen}{rgb}{0,0.5,0}

\definecolor{Code}{rgb}{0,0,0}
\definecolor{Decorators}{rgb}{0.5,0.5,0.5}
\definecolor{Numbers}{rgb}{0.5,0,0}
\definecolor{MatchingBrackets}{rgb}{0.25,0.5,0.5}
\definecolor{Keywords}{rgb}{0,0,1}
\definecolor{self}{rgb}{0,0,0}
\definecolor{Strings}{rgb}{0,0.33,0}
\definecolor{Comments}{rgb}{0,0.33,1}
\definecolor{Comments}{rgb}{0.5,0.5,0.5}
\definecolor{Backquotes}{rgb}{0,0,0}
\definecolor{Classname}{rgb}{0,0,0}
\definecolor{FunctionName}{rgb}{0,0,0}
\definecolor{Operators}{rgb}{0,0,0}
\definecolor{Background}{rgb}{1,1,1}
\usepackage{calc}

\begin{document}
\title{Don't Tell Me What Is, Tell Me What Ought To Be! 
Learning Effective Changes for  Software Projects
\\[-0.9cm]}
\author{Rahul Krishna\\
	Comptuer Science, North Carolina State University, USA\\
	i.m.ralk@gmail.com}

\maketitle

\begin{abstract}

The primary motivation of much of software analytics is decision making. How to make these decisions? Should one make decisions based on lessons that arise from within a particular project? Or should one generate these decisions from across multiple projects? This work is an attempt to answer these questions. Our work was motivated by a realization that much of the current generation software analytics tools focus primarily on prediction. Indeed prediction is a useful task, but it is usually followed by ``planning'' about what actions need to be taken. This research seeks to address the planning task by seeking methods that support actionable analytics that offer clear guidance on \textit{what to do}. Specifically, we propose XTREE and BELLTREE algorithms for generating a set of actionable plans within and across projects. Each of these plans, if followed will improve the quality of the software project.

\end{abstract}


\section{Introduction}
\label{sect:intro}

Over the past decade, advances in AI have enabled a widespread use of data analytics in software engineering. For example, we can now estimate how long it would take to integrate the new code~\cite{czer11}, where bugs are most likely to occur~\cite{Menzies2007a}, or amount of effort it will take to develop a software package~\cite{turhan11}, etc. Despite these successes, there are two primary operational shortcomings with many software analytic tools: (a) conclusion instability as a result of constant influx of new data; and (b) lack of insightful analytics.

In several applications where local data is scarce, researchers use transfer learning. They report that the use of data from other projects can yield comparable predictors to just using local data~\cite{peters15}. However, new projects are constantly being created. Rahman et al.~\cite{rahman12} caution that if quality predictors are always being updated based on the specifics of new data, then those new predictors may suffer from over-fitting. Such over-fitted models are ``brittle'' in the sense that they can undergo constant changes whenever new data arrives and lead to unstable conclusions. Conclusion instability is unsettling for software project managers struggling to find general policies.
We require methods to support managers, who seek stability in their conclusions, while also allowing new projects to take full benefit from data arriving from all the other projects. Our research~\cite{krishna16} has offered strong evidence that organizations can declare some prior project as  the {\em ``bellwether''}\footnote{According to the Oxford English Dictionary, the ``bellwether'' is the leading sheep of a flock, with a bell on its neck.} that can then offer predictions that generalize across $N$ other projects.

In addition to unstable conclusions, business users also lament that most software analytics tools, ``Tell us what \textit{is}. But they don't tell us \textit{what to do}''. A concern that was also raised by several researchers at a recent workshop on ``Actionable Analytics'' at 2015 IEEE conference on Automated Software Engineering~\cite{hihn15}. For example, most software analytics tools in the area of detecting software defects are mostly \textit{prediction} algorithms such as Support Vector Machines, Naive Bayes, Logistic Regression, Decision Trees, etc~\cite{scikit-learn}. These prediction algorithms report what combinations of software project features predict for the number of defects. But this is different task to \textit{planning}, which answers a more pressing question: what to {\em change} in order to {\em reduce} these defects. Accordingly, in this research, we seek tools that offer clear guidance on what to do in a specific project.

The tool assessed in this paper is the XTREE \textit{planning} tool~\cite{krishna17a}. XTREE employs a $cluster+contrast$ approach to planning where it (a) \textit{Clusters} different parts of the software project based on a quality measure (e.g. the number of defects); (b) Reports the \textit{contrast sets} between neighboring clusters. Each of these contrast sets represent the difference between these clusters and they can  be interpreted as plans, i.e., 
\begin{itemize}
	    \item If a current project falls into cluster $C_1$,
	    \item Some neighboring cluster $C_2$ has better quality.
	    \item Then the difference {\em $\Delta=$ $C_2$ - $C_1$} is a {\em plan} for changing a  project such that it \textit{might} have   higher quality.
\end{itemize}

XTREE uses data from within a software project to generate plans. But, in several cases local data may not readily available. To overcome this limitation, we incorporate our findings from bellwethers to extend XTREE to use the bellwether projects. We call this tool BELLTREE and we show that it can be used to generate \textit{stable} plans for cross-company planning. 

\section{Contributions of this work}
\label{sect:contributions}

\noindent\textit{1. New kinds of software analytics techniques:} This research introduces the notion of planning in software engineering. In addition to showing that planning in effective in a within-project setting~\cite{krishna17a}, we also show that with bellwethers~\cite{krishna16}, plans can be generated for cross-project problems with encouraging results. 
This is a unique approach that combines our efforts to address the problems highlighted in \tion{intro}.  

\noindent\textit{2. Compelling results of planning:} Our results have established that planning is quite successful in producing actions that can reduce the number of defects. In~\fig{rq1_1}, we show that planning can reduce defects by more than 40\% in 3 out of the 4 datasets studied here ($>$80\% in the certain cases).

\noindent\textit{3. Evidence of generality of bellwethers:}
The more the bellwether effect is explored, the more we learn about its broad applicability. Originally, we explored this just in the context of defect prediction~\cite{krishna16}, it has now been shown to work also in effort estimation, predicting when issues will close, and detecting code smells~\cite{krishna17b}. Our preliminary results reported in this work show that bellwethers can also be used for cross-project planning with the use of BELLTREE. This is an important result of much significance since, where bellwethers occur, reasoning about multiple software projects becomes a simple matter of discovering bellwethers (see~\cite{krishna16}).

\noindent\textit{4. Replication Package:} For readers this work who wish to replicate our findings, we have made available a replication package at \url{https://git.io/v7c9k}.

\section{Related Work}

Planning  has been a subject of much research in artificial intelligence. Here, planning usually refers to generating a sequence of actions that enables an \textit{agent} to achieve a specific \textit{goal}~\cite{norvig}. This can be achieved by classical search-based problem solving  approaches or logical planning agents. Such planning tasks now play a significant role in a variety of demanding applications, ranging from controlling space vehicles and robots to playing the game of bridge~\cite{ghallab04}. Some of the most common planning paradigms include: (a) classical planning~\cite{wooldridge95}; (b) probabilistic planning~\cite{Bel, altman99, guo2009}; and (c) preference-based planning~\cite{son06, baier09}. 

Existence of a model precludes the use of each of these planning approaches. This is a limitation of all these planning approaches since not every domain has a reliable model. In software engineering, the planning problem translates to proposing changes to software artifacts. Solving this has been undertaken via the use of some search-based software engineering techniques~\cite{Harman2009}. Examples of algorithms include SWAY, NSGA-II, MOEA/D, etc.~\cite{nair2016accidental,deb00a,zit02}.

These search-based software engineering techniques require access to some trustworthy models that can be used to explore novel solutions. In some software engineering domains there is ready access to such models which can offer assessment of newly generated plans. Examples of such domains within software engineering include automated program repair~\cite{Weimer2009, LeGoues2015}, software product line management~\cite{sayyad13, henard15}, etc.

However, not all domains come with ready-to-use models. For example, consider software defect prediction and all the intricate issues that may lead to defects in a product. A model that includes {\em all} those potential issues would be very large and complex. Further, the empirical data required to validate any/all parts of that model can be hard to find. Also, even when there is an existing model, they can require constant  maintenance lest they become out-dated. These problems are the key motivations for us to look for alternate methods for planning that can be automatically updated with new data without a need for comprehensive models.

In summary, for domains with readily accessible models, we recommend
the tools widely used in the search-based
software engineering community such as SWAY, NSGA-II, MOEA/D, etc. In cases where this is not an option, we propose the use of data mining approaches to create a quasi-model of the domain 
and make of use observable states from this data to generate an estimation of the model. Our preferred tools in this paper XTREE and BELLTREE take this approach and as presented elsewhere in this paper, these methodologies have very encouraging results.

\section{Planning in Software Engineering}

\subsection{What is planning?}
We distinguish planning from prediction for software quality as follows: 
Quality prediction points to the likelihood of defects. Predictors take the form:
$$
    out = f(in)    
$$
where $in$ contains many independent features and out contains some measure of
how many defects are present. For software analytics, the function $f$ is learned via data mining (with static code attributes for instance). Contrary to this, quality planning generates a concrete set of actions that can be taken (as precautionary measures) to significantly reduce the likelihood of defects occurring in the future. For a formal definition of plans, consider a test example $Z$, planners
proposes a plan $D$ to adjust attribute $Z_j$ as follows:
{\small\[
\forall \delta_j \in \Delta :  Z_j =  
\begin{cases}
     Z_j + \delta_j& \text{if $Z_j$ is numeric}\\
    \delta_j              & \text{otherwise}
\end{cases}
\]}
With this, to (say) simplify a large bug-prone method, our planners
might suggest to a developer to reduce its size (i.e. refactor that
code by splitting it simpler functions).


\subsection{XTREE}
\label{sect:xtree}
XTREE builds a {\em supervised} decision tree and then generates
plans by contrasting the differences between two branches:
(1)~branch where you are; (2)~branch to where you want to be.

The specifics of the algorithm used to divide the data and construct the decision tree were presented in greater detail in our previous work~\cite{krishna17a}.
Next, XTREE builds plans from the branches of the decision tree by asking the following three questions for each test case (the last of which returns the plan):
\be
\item
Which {\em current} branch does a test instance fall in?
\item Which {\em desired} branch would we want to move to?
\item What are the {\em deltas} between current and desired? 
\ee
\begin{figure}
\includegraphics[width=\linewidth]{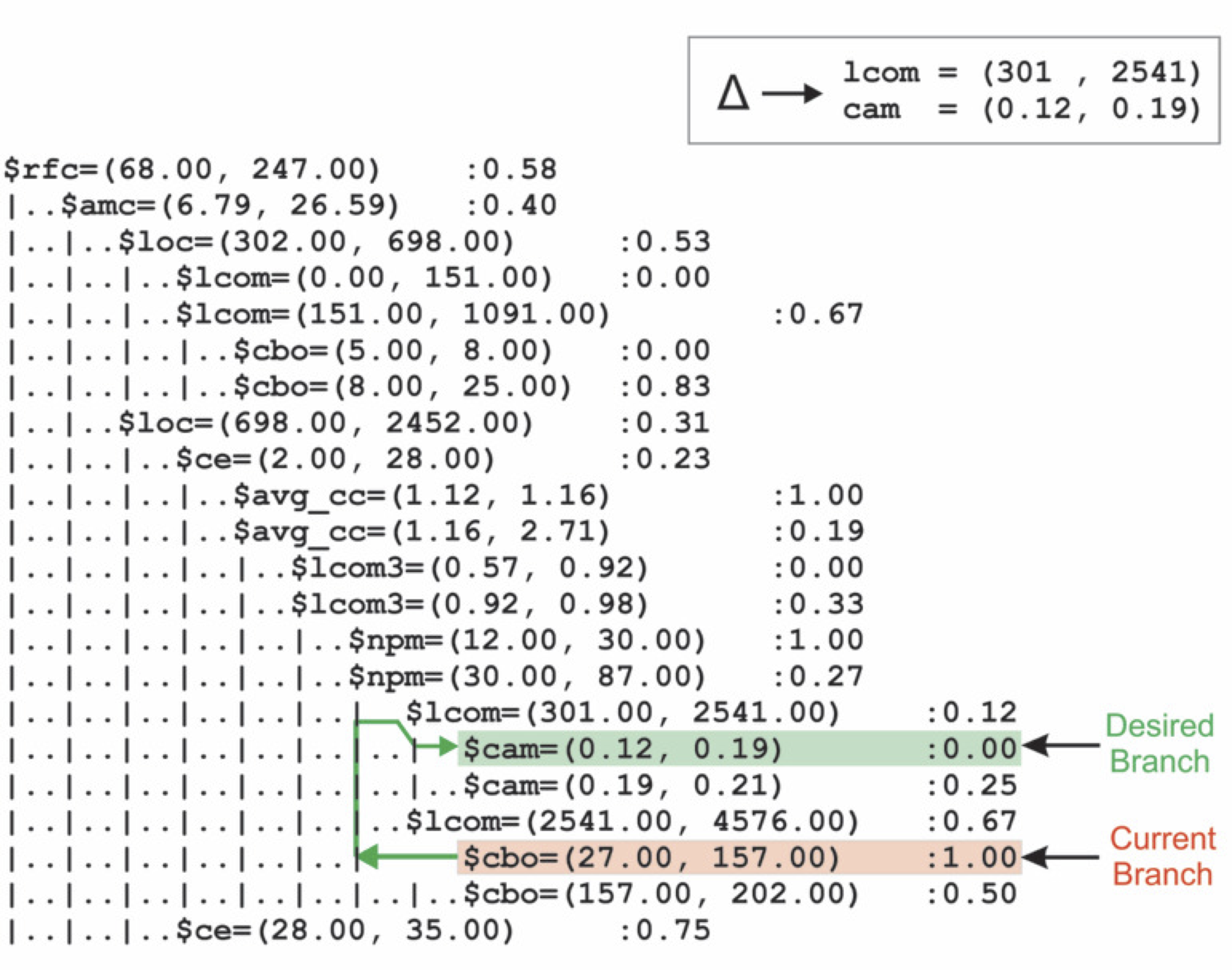}
\caption{Generating thresholds using XTREE.}
\label{fig:xtree}
\end{figure}

As a motivating example, consider~\fig{xtree} with XTREE constructed with training data consisting of OO code metrics~\cite{ck} and associated defect counts. 
A defective test case with the same code metrics is passed into the 
tree and evaluated down the tree to a leaf node with a defect probability of 1.0 (see the \textcolor{orange}{{\bf orange}} line in \fig{xtree}).
XTREE then looks for a nearby leaf node with a lower defect
probability (see the \textcolor{green}{{green}} line in \fig{xtree}). XTREE then evaluates the differences (of $deltas$) between
\textcolor{green}{{green}} and \textcolor{orange}{{\bf orange}}.
These \textit{deltas} represent the threshold ranges\footnote{Thresholds are denoted by $[low,high)$ ranges for each OO metric} that represent the plans to reduce the defects.

\subsection{BELLTREE}
BELLTREE is structurally similar to XTREE. It differs in the source of data used for analytics. While XTREE uses data from within the project, BELLTREE first starts by looking for the bellwether dataset. To do this, we employ the strategy discussed in our previous work~\cite{krishna16}. This helps in identifying a bellwether dataset. Once the bellwethers are discovered, we construct a supervised decision tree similar to XTREE. Plans are generated by using the same procedure as \tion{xtree}. Note that the use of bellwethers enables BELLTREE to leverage data from across different projects. This presents a novel extension to XTREE.

\section{Research Questions}

\noindent\textit{RQ1. How prevalent are bellwethers?}
It is important to establish the prevalence of bellwethers first as this determines if it is possible to learn plans from the bellwether data. If bellwethers occur infrequently, we cannot rely on them for planning. We have initially shown that bellwethers are prevalent in defect prediction~\cite{krishna16}. Further evidence was seen in ~\cite{krishna17b}, where we explored three additional sub-domains within software engineering namely, defect prediction, effort estimation, issue lifetime estimation, and detection of code smells. In a result consistent with bellwethers being \textit{very} prevalent, we found that all these domains have a bellwether dataset.

\noindent\textit{RQ2. Does within-project planning with XTREE offer significant improvements in reducing defects?}
This research question seeks to establish if our preferred planning tool (XTREE) is effective in generating actionable plans in a within-project setting. Our initial findings showed that XTREE was indeed an effective planner that can generate plans that are also succinct and stable. Further, these plans are not subject to conjunctive fallacy~\cite{krishna17a}.

\noindent\textit{RQ3. Does  cross-project  planning  with  BELLTREE offer significant improvements in reducing defects?}
Having established the prevalence of bellwether datasets and the efficacy of planning with XTREE, here we ask if it is possible for us to transfer plans across projects using the bellwether data and XTREE (referred to as BELLTREE). Our preliminary results are very encouraging. We show that BELLTREE can be a very effective cross-project planner.

\noindent\textit{RQ4. Are cross-project  plans  any  better  than within project plans?}
This research question assesses the quality of plans obtained using XTREE and BELLTREE. This is important because within-project data is not always available (especially if a project is in it's early stage of development) and it may be useful to look to other similar projects for planning. Our preliminary results have suggested that the effectiveness of plans generated from within project data and XTREE is statistically comparable to plans derived with cross-project data and BELLTREE. Thus, when project specific data is not available, one may use cross-project data to derive plans.

\section{Evaluating Plans}

To evaluate plans, we propose the use of a \textit{verification oracle}~\cite{krishna17a}. Oracles have been commonly used by several SE researchers such as Cheng et al~\cite{Cheng10}, O'keefe et al.~\cite{Okeeffe08}, Mkaouer et al.~\cite{Mkaouer14}. They use an oracle that is learned separately from the planner. The verification oracle assesses
how defective the code is before and after some
code changes.
For their oracle,
Cheng, O'Keefe, Moghadam and  Mkaouer et al. use the QMOOD
quality model~\cite{Bansiya02}.
A shortcoming of QMOOD
is that quality models learned from other projects
may perform poorly when applied to new projects~\cite{localvsglobal}.

Hence, for this study, we  eschew
older quality models like QMOOD. Instead, we use
Random Forests~\cite{Breiman2001} to learn defect predictors
from OO code metrics. 
Unlike QMOOD, the predictors
are specific to the project. Additionally, classifiers such as Random Forest have shown to be very efficient in detecting bugs~\cite{fu}.

For planning and construction of a verification oracle, we divide the
project data into two parts the \textit{train set} and the \textit{test test}.
The train set could either be data that is available locally within a project, or it could be data from the bellwether dataset. We further partition the train set to build both a {\em planner} and a {\em verification oracle}. It is important to note that: 
\begin{quote}
{\em The verification oracle should be built with completely different data to the planner.}
\end{quote}

After constructing the planner and verification oracle, we (1)~deploy the {planner} to recommend plans; (2)~alter the {\em test} data according to these plans;
then (3)~apply the {verification oracle} to the altered data to estimate defects; then (3)~Compute the percent improvement, denoted by the following equation:
\begin{equation}
\small
	\label{eq:diff}
	R=(1-\frac{\mathit{after}}{\mathit{before}})\times100\%
\end{equation}
The value of the measure $R$ has the following properties:
\bi
\item If $R = 0\%$, this means  ``no change from baseline''; 
\item If $R > 0\%$, this indicates ``improvement over the baseline'';
\item If $R < 0\%$, this indicates ``optimization failure''.
\ei

Ideally, an effective planner should have an improvement of $R>0$, where larger values indicate better performance. 

\section{Current State and Future Work}
\begin{figure}[tbp]
\scriptsize
\begin{minipage}{0.99\linewidth}{}
{\scriptsize{\hspace{3.5cm}\underline{Observed Improvements (from \eq{diff})}}\vspace{2mm}}
\resizebox{\linewidth}{!}{\textbf{Ant}~~~~~~~~ \begin{tabular}{{l@{~~~~}l@{~~~~}r@{~~~~}r@{}c@{~~~}r}}
\arrayrulecolor{lightgray}
\rowcolor{lightgray}\textbf{Rank} & \textbf{Treatment} & \textbf{Median} & \textbf{IQR~~~} & \bigstrut\\
1 & XTREE &    44.0  &  6.0 & \quart{69}{10}{73}{1} \bigstrut\\
\hline 2 & BELLTREE &    39.0  &  16.0 & \quart{48}{26}{64}{1}   \bigstrut\\
\end{tabular}}\bigstrut\\

\resizebox{\linewidth}{!}{
\textbf{Poi}~~~~~~~~ \begin{tabular}{{l@{~~~}l@{~~~}r@{~~~}r@{~~~}c}}
\arrayrulecolor{lightgray}
\rowcolor{lightgray}\textbf{Rank} & \textbf{Treatment} & \textbf{Median} & \textbf{IQR~~~} & \bigstrut\\
  1 & XTREE &    84.0  &  6.0 & \quart{74}{5}{78}{0} \bigstrut\\
  1 & BELLTREE &    83.0  &  3.0 & \quart{74}{3}{77}{0} \bigstrut\\
\end{tabular}}\bigstrut\\

\resizebox{\linewidth}{!}{\textbf{Ivy}~~~~~~~~ \begin{tabular}{{l@{~~~}l@{~~~}r@{~~~}r@{~~~}c}}
\arrayrulecolor{lightgray}
\rowcolor{lightgray}\textbf{Rank} & \textbf{Treatment} & \textbf{Median} & \textbf{IQR~~~} & \bigstrut\\
  1 & BELLTREE &    25.0  &  12.0 & \quart{54}{25}{54}{2} \bigstrut\\
  1 & XTREE &    24.0  &  12.0 & \quart{45}{26}{51}{2} \bigstrut\\
\end{tabular}}\bigstrut\\

\resizebox{\linewidth}{!}{ \textbf{Jedit}~~~~~ \begin{tabular}{{l@{~~~}l@{~~~~}r@{~~~~}r@{~~}c@{}r}}
\arrayrulecolor{lightgray}
\rowcolor{lightgray}\textbf{Rank} & \textbf{Treatment} & \textbf{Median} & \textbf{IQR~~~} & \bigstrut\\
  1 & XTREE &    63.0  &  2.0 & \quart{77}{2}{78}{1} \bigstrut\\
\hline   2 & BELLTREE &    60.0  &  9.0 & \quart{68}{11}{74}{1} \bigstrut\\
\end{tabular}}\bigstrut\\
\end{minipage}
\caption{Results comparing XTREE trained on local datasets and BELLTREE.  Results from 30 repeats.
Values come from Eq. 1.
Values near 0
imply no improvement,
{\em Larger} median values are {\em better}. }
\label{fig:rq1_1}
\end{figure}
As mentioned earlier in the paper, this work represents our efforts to address to key issues in modern software analytics: (a) conclusion instability; and (b) generating insightful analytics. To this end, we undertook two concurrent research efforts to address each of these issues. 

While attempting to stabilize the pace of conclusion change, we discovered the bellwether effect~\cite{krishna16}. Our results provided evidence that it is possible to \textit{slow} the pace of conclusion change in software analytics (for defect prediction models) using bellwethers. Further exploration demonstrated that the so called \textit{bellwether effect} is quite prevalent in several sub-domains of software engineering such as code-smell detection, effort estimation, and estimation of issue lifetimes~\cite{krishna17b}.

In order to generate actionable analytics for software engineering, we developed the XTREE planner~\cite{krishna17a}. Initial motivation for XTREE was to address the varied opinions in literature on how best to undertake code reorganization so as to reduce bad smells. We showed that by leveraging historical logs of data, planners such as XTREE can offer actionable recommendations on how to undertake code reorganization in order to reduce defects in code. Further, we showed that in addition to generating effective plans, XTREE recommends of far fewer changes. Thus making it a better framework for critiquing and rejecting many of the code reorganizations.


The initial version of XTREE was limited to using data from within a project to generate plans. This paper represents our initial attempts to transfer plans from across other projects to a test project. For this purpose, we developed BELLTREE. It uses the same framework as XTREE but uses bellwethers as the source of data for planning.
Our results comparing BELLTREE with XTREE on a set of open source java projects is shown in \fig{rq1_1}. In two of the four datasets, we note that BELLTREE performed just as well as XTREE and two other cases XTREE outperformed BELLTREE (but not by a significant amount). Our initial finding is that if within-project data from previous releases are available, we may use XTREE. If not, using bellwethers would be a reasonable alternative.


Our initial results of using BELLTREE are encouraging and deserves much further exploration. Starting early this summer, we have deployed an enhanced version of XTREE on-site in conjunction with our industrial partners with the following goals: (1) Qualitatively validate the usefulness of the plans; (2) Establish the receptiveness of developers actively using our tool; and (3) Solicit developers' feedback on usefulness of plans generated by XTREE.  
\balance
\bibliographystyle{IEEEtran}
\bibliography{manuscript}

\end{document}